\documentclass[oribibl]{llncs}

\makeatother

\usepackage[colorlinks=false]{hyperref}

\usepackage{enumitem}
\newlist{inlinelist}{enumerate*}{1}
\setlist*[inlinelist,1]{%
  label=(\roman*),
}
\usepackage{subfigure}
\usepackage{acronym}
\usepackage{xcolor}
\usepackage[numbers]{natbib}
\usepackage{graphicx}

\acrodef{LLM}{large language models}
\acrodef{CoT}{chain of thought}

\title{Interactions with Generative Information Retrieval Systems}

\author{Mohammad Aliannejadi\inst{1} \and
Jacek Gwizdka\inst{2} \and Hamed Zamani\inst{3}}
\institute{University of Amsterdam, The Netherlands\\\email{m.aliannejadi@uva.nl} \and
University of Texas at Austin, United States\\\email{jacekg@utexas.edu} \and
University of Massachusetts Amherst\\\email{zamani@cs.umass.edu}
\\~\\~\\
Draft of a chapter intended to appear in a forthcoming book on generative information retrieval, co-edited by Chirag Shah and Ryen White.
}

\begin{document}

\maketitle

\section{Introduction}

At its core, information access and seeking is an interactive process. In existing search engines, interactions are limited to a few pre-defined actions, such as ``requery'', ``click on a document'', ``scrolling up/down'', ``going to the next result page'', ``leaving the search engine'', etc. A major benefit of moving towards generative IR systems is enabling users with a richer expression of information need and feedback and free-form interactions in natural language and beyond. In other words, the actions users take are no longer limited by the clickable links and buttons available on the search engine result page and users can express themselves freely through natural language. This can go even beyond natural language, through images, videos, gestures, and sensors using multi-modal generative IR systems. This chapter briefly discusses the role of \emph{interaction} in generative IR systems. We will first discuss different ways users can express their information needs by interacting with generative IR systems (Section~\ref{infoneeds}). We then explain how users can provide explicit or implicit feedback to generative IR systems and how they can consume such feedback (Section~\ref{feedback}). Next, we will cover how users interactively can refine retrieval results (Section~\ref{refinement}). We will expand upon mixed-initiative interactions and discuss clarification and preference elicitation in more detail (Section~\ref{clarification}). We then discuss proactive generative IR systems, including context-aware recommendation, following up past conversations, contributing to multi-party conversations, and feedback requests (Section~\ref{proactive}). Providing explanation is another interaction type that we briefly discuss in this chapter (Section~\ref{explanation}). We will also briefly describe multi-modal interactions in generative information retrieval (Section~\ref{multimodal}). Finally, we describe emerging frameworks and solutions for user interfaces with generative AI systems (Section \ref{interface}). We conclude with a question: Will the myriad interaction possibilities afforded by generative AI systems be embraced by a broad user base, or will they remain merely a research curiosity?

\section{Expressing Information Needs} %
\label{infoneeds}

An information need is what prompts users to seek information through various means, such as asking others, consulting printed resources, other media, or searching online. It arises from the awareness of a gap in a user’s knowledge or understanding, necessitating the acquisition of information to bridge that gap
~\cite{ belkin1980Anomalousstatesknowledge, dervin1986information}. Bridging the gap helps to fulfill a specific purpose or goal, which is typically driven by a work task ~\cite{bystrom2005Conceptualframeworktasksb}.

Prompt-based interactions with \acp{LLM}, and more broadly, multi-modal interactions with \acp{LLM}-based systems, provide an opportunity to fundamentally rethink the processes of searching for, finding, and using information, and how to support these activities. This fresh perspective has the potential to significantly transform user experience by enhancing how users express their information needs and achieve their goals.

We will frame our considerations using the information need model proposed by Robert Taylor in the 1960s~\cite{taylor1962ProcessAskingQuestions, taylor1968QuestionNegotiationInformationSeeking}. Taylor identified four levels of information need, each helping us to understand how users formulate questions in their minds, how they articulate them, and how they interact with information systems. The four levels of information need are: (1) \textbf{Visceral Need}: An inexpressible, unformulated need, felt as a vague sense of dissatisfaction, (2) \textbf{Conscious Need}: The user is aware of the need but cannot fully articulate it, (3) \textbf{Formalized Need}: The need can be clearly expressed and defined, (4) \textbf{Compromised Need}: The articulated need, as presented to an information system, often simplified or altered to fit the system's capabilities.

Traditional search systems typically support levels 3 and 4, but not 1 and 2. We believe that \acp{LLM}-based information access systems have the potential to support all four levels. Therefore, we use these four levels to structure our speculative list of ways users could be assisted in their interactions with generative AI. We will draw, in part, on well-known information seeking models
\cite{marchionini1995InformationSeekingElectronic, kuhlthau1991searchprocessInformation}. 

Support for \textbf{Visceral Need}:  
(1) \textit{exploratory interactions}: provide users with broad, exploratory dialogue that might help users \textit{clarify} their thoughts; suggest related topics to help users better understand and articulate their needs. This is an example of \nameref{clarification} which we describe in Section \ref{clarification}. 
(2) \textit{prompt suggestions}: offer prompt suggestions or follow-up questions to guide users towards more specific questions. This is an example of \nameref{proactive} which we describe in Section \ref{proactive}. 

Support for \textbf{Conscious Need}:
(1) \textit{partial expression of needs}: accept partially formed questions or statements of need;
(2) \textit{proactive support for refinement}: generate relevant information that helps users \textit{refine} their understanding of what they're looking for; 
(3) \textit{guided conversations}: engage in a dialogue to help users articulate their needs more precisely. We describe such approaches in more detail in \nameref{refinement} (Section \ref{refinement}) and \nameref{proactive} (Section \ref{proactive}). 

Support for \textbf{Formalized Need}:
(1) \textit{direct queries}: respond directly to well-formulated questions with relevant information; 
(2) \textit{structured responses}: provide detailed, structured responses that address specific aspects of the user's need;
(3) \textit{advanced features}: offer options (e.g., filters) for further exploration or \textit{clarification} based on the formalized need.

Finally, support for \textbf{Compromised Need}:
(1) \textit{flexibility of syntax}: offer flexibility to allow for iterative refinement of queries without strict syntax requirements;
(2) \textit{flexibility of language}: interpret and respond to a wide range of query formats, reducing the need for users to adapt their language significantly;
(3) \textit{feedback loop}: offer feedback on questions, suggesting modifications or alternative phrasings to better match the user's needs and system’s capabilities. We describe such approaches in more detail in \nameref{feedback} (Section \ref{feedback}).

Overall, the key advantages of \acp{LLM} in assisting users at all four levels of information need are:
\begin{itemize}
    \item \textbf{Natural Language Processing}: \acp{LLM} can understand and respond to queries expressed in natural language, making them accessible even at the visceral and conscious need levels.

    \item \textbf{Contextual Understanding}: Advanced \acp{LLM} can maintain context over multiple interactions, allowing for a more nuanced exploration of information needs.
    
    \item \textbf{Broad Knowledge Base}: \acp{LLM} draw upon a vast range of information, potentially addressing needs across various domains and levels of specificity.
    
    \item \textbf{Adaptive Responses}: \acp{LLM} can tailor their responses based on the perceived level of the user's information need, understanding and responding to both simple and complex questions and providing more or less detail as appropriate.
    
    \item \textbf{Iterative Refinement}: The conversational nature of \acp{LLM} interactions allows users to refine their queries progressively, moving from visceral to formalized needs through dialogue

    \item \textbf{Enhanced Expressiveness}: Prompt-based interactions allow users to express their needs in more nuanced and detailed ways. Users can specify the format, tone, and depth of the information they seek, which can lead to more tailored and useful outputs. For instance, users can request summaries, detailed explanations, comparisons, or creative content, depending on their needs.
\end{itemize}

However, it's important to note that while \acp{LLM} offer powerful capabilities in addressing information needs across Taylor's levels, they also have limitations. They may sometimes provide plausible-sounding but incorrect information, lack true understanding of context beyond the immediate conversation, and cannot replace the critical thinking and expertise of human information professionals in complex scenarios.

While \acp{LLM} can offer enhanced capabilities for expressing information needs, they also introduce new challenges. Such as \textit{Capability Gap}: users may struggle to formulate their intentions clearly and effectively, leading to a gap between what they want and what the \acp{LLM} provides.
\textit{Instruction Gap}: users need to learn how to craft effective prompts, which can involve understanding the \acp{LLM}'s capabilities and limitations.
\textit{Evaluation of Outputs}: users must critically evaluate the \acp{LLM}'s responses for accuracy and relevance, as \acp{LLM} can sometimes generate incorrect or misleading information. 
A recent paper introduced these three gaps and termed them  collectively the \textit{''Gulf of Envisioning''} ~\cite{ subramonyam2024BridgingGulfEnvisioning}.

In the following sections we address selected aspects of user-\acp{LLM}-based-system-interactions, \ref{feedback} \nameref{feedback} \ref{refinement}, \nameref{refinement} \ref{clarification} \nameref{clarification}, \ref{proactive} \nameref{proactive}, \ref{explanation} \nameref{explanation}, and \ref{multimodal} \nameref{multimodal}. In Section \ref{interface}, \nameref{interface}, we discuss recent user interface frameworks and solutions.

\section{Proactive Feedback} %
\label{feedback}

Recent developments in large language models have paved the path towards complex interactions between the user and the system. Generative IR models are able to satisfy user's information needs in multiple interaction turns. Among many possibilities, this enables users to provide feedback to the system. Feedback can be provided when is explicitly requested by the system, for example in the form of clarifying questions or preference elicitation \cite{Aliannejadi:2019:asking,Zamani:2020:generating,Rao:2018:learning,Radlinski:2019:coached}. Section~\ref{clarification} discusses these aspects in more detail. Feedback can be also requested for assessing the quality of the system at the end or in the middle of a conversation. For instance, Amazon's Alexa Prize Challenge \cite{alexaprize} has sought explicit rating feedback from users upon the completion of the conversation. Zamani et al.~\cite{CISbook} introduce the possibility of improving this simple approach by asking context-aware questions for feedback and making natural language interactions within the conversation.

Feedback can be provided proactively by the user, which is the focus of this section. Perhaps the simplest type of feedback that users provide can be in the form of \emph{repeating or reformulating the user's need in the same search session}. If detected, this often means that the user's need has not been addressed yet. Besides such simple scenarios, users may provide \emph{explicit positive or negative feedback}. Explicit positive feedback are often easier to identify and interpret. They are often in the form of appreciation and hold a positive sentiment. Explicit negative feedback, on the other hand, is more challenging, more diverse, and perhaps more important for system designers as they help the system to improve and identify its limitations. Pointing out what parts of the system's response is inaccurate, why it is does not satisfy the user's needs, or expressing frustration and disappointment are examples of explicit negative feedback. Current state-of-the-art technologies often cannot successfully take advantage of explicit negative feedback and often limit themselves to acknowledging the system's limitations and apologizing from the users. There are huge potentials in successfully comprehending negative feedback from users.

In generative IR systems, grounding as relevance feedback is also relevant to the concept of explicit feedback. Trippas et al.~\cite{Trippas:2020:SpokenConversation} define grounding as discourse for the creation of mutual knowledge and beliefs. Examples include providing indirect feedback by reciting their interpretation of the results. This process can potentially enable CIS systems to better understand a user’s awareness of the results, background knowledge, or information need.

We would like to highlight the potentials in providing implicit feedback as well. Progress in commercial (web) search engines is in debt to large-scale implicit feedback collected from user interactions, such as clicks, skipped results, dwell time, and cursor (mouse) movement. Implicit feedback in generative IR systems is more challenging, because it is more likely to deal with abandonment in each session. This means that users may leave the system as they receive the answer they want without providing any positive feedback. Alternatively, they may leave the system as they lose hope in getting the right answer from the system. Besides abandonment, changing topics and asking follow-up questions can be interpreted as an implicit feedback signal in generative IR. Interpreting these user behaviors is essential in improving generative IR systems.

Research in understanding and modeling implicit (negative) feedback is relatively sparse and future technologies can greatly benefit from further research in this space.

\section{Result Refinement} %
\label{refinement}

\subsection{An Overview of Result Refinement}

\begin{figure}
    \centering
    \subfigure[]{
        \includegraphics[width=0.45\linewidth]{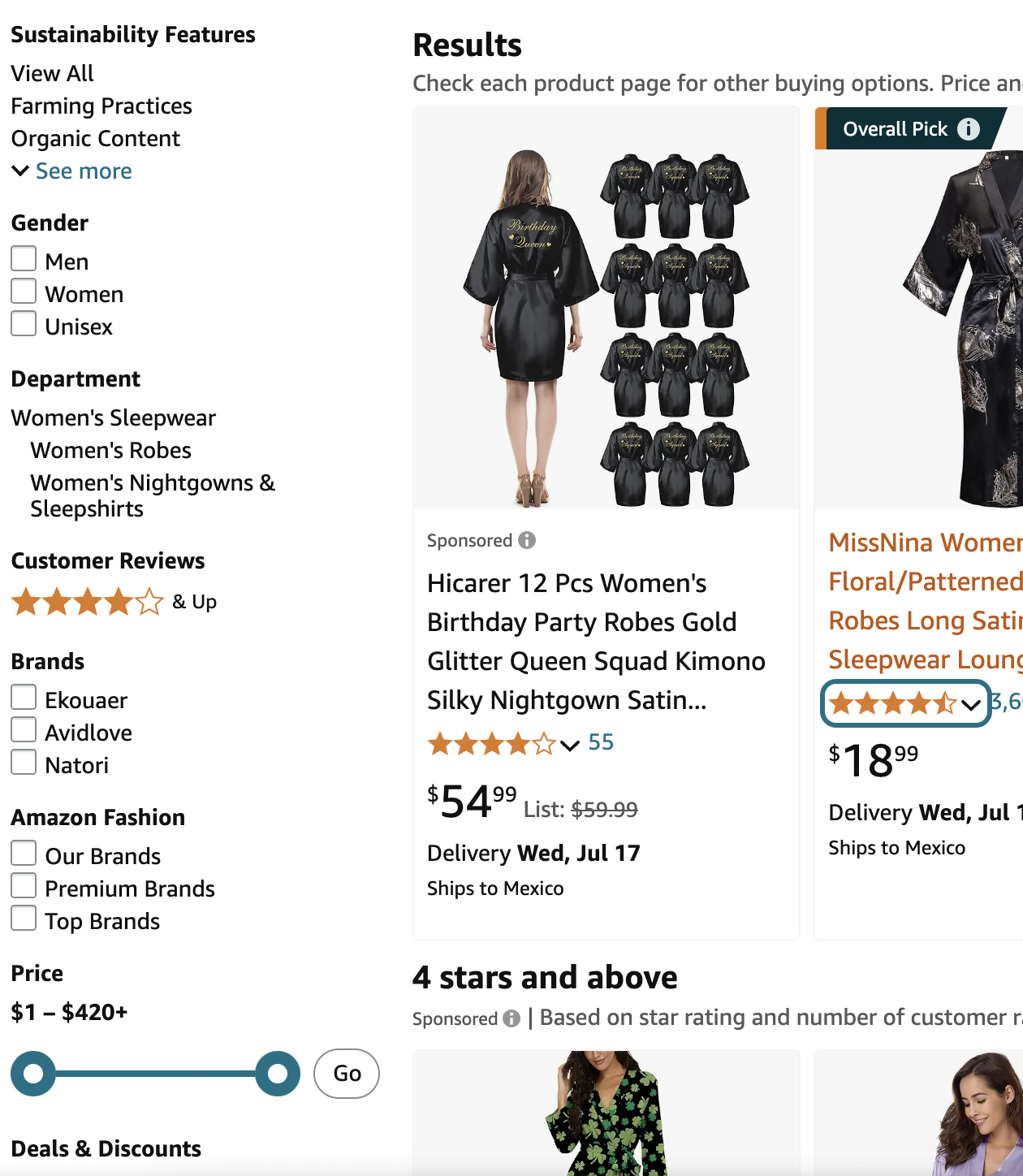}
    }
    \subfigure[]{
        \includegraphics[width=0.45\linewidth]{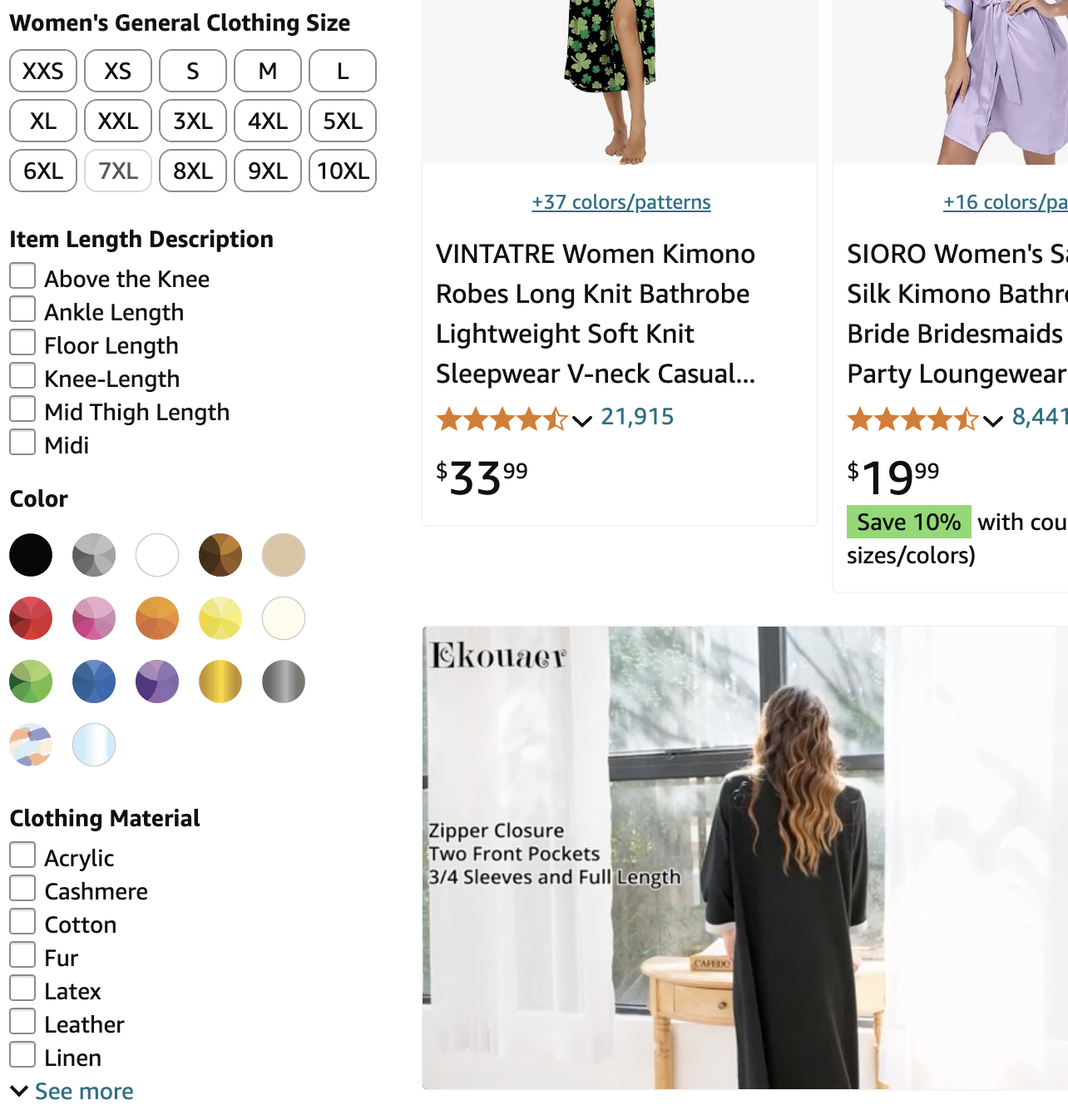}
    }
    \caption{Examples of search result refinement from Amazon.com. The refinement panes on the left help users browse through the search results.}
    \label{fig:refinement}
\end{figure}

Result refinement is relatively understudied, compared to other modes of interaction in generative IR. Search result refinement has a long history of research in IR, especially in areas such as information filtering (e.g., recommender systems) where users access semi-structured information~\citep{Cui:2021:disentangled}. Figure~\ref{fig:refinement} shows an example search result page from Amazon.com, where users are able to select certain attributes of the items (e.g., size) in the catalog to narrow down the results being presented to them. Search result refinement for semi-structured data is a relatively trivial task, as the refinement pane usually concerns the most important attributes of the items, given the query and the top item list. In the preference-based search literature, example-critiquing approaches have been explored~\cite{Viappiani:2006:preference}, where the model suggests examples to the user, and with the user's feedback, it then models the user's preference. In conversational recommender systems, a similar approach is taken as part of the preference elicitation process~\cite{Kostric:2024:generating}. In this process, the conversational system starts the conversation by asking the user's opinion about movies, aiming to optimize the decision space. A similar approach is taken in conversational product recommendation~\cite{Zou:2019:learning,Zou:2020:towards}. In these works, the high-level idea is to extract important attributes from user reviews of products and model a probabilistic decision space. Then the conversational system takes a greedy approach in which, at every step, it aims to ask about an item attribute that minimizes the uncertainty of the decision space. 
Search result refinement is more challenging in web search, where the system deals with unstructured data. One of the earliest, simplest, and yet most effective ways is using vertical in the search result page~\cite{Arguello:2011:learning}. Search result verticals divide the search results based on very high-level categories, such as images, videos, and news. Even though, very high level, it still can be considered as a naïve approach to refinement, as it approaches the user information from the result type. In most cases, the same user query can be satisfied with different modalities, which turns out to be one of the most important aspects of search, hence major commercial search engines still employ this approach.  
Finally, some early approaches tried to diversify, but also refine search results based on automatically extracted information facets. Faceted search~\cite{Tunkelang:2009:faceted} provides a means of navigation through topic facets for the users, enabling them to narrow down their information needs, as well as the search space. These early systems mainly relied on automatic facet extractors~\cite{Kong:2013:extracting}. 

\subsection{Technical Challenges}
In the generative era, result refinement faces both algorithmic and interactive challenges. 

\textbf{Algorithmic challenges.}
As the items or documents are being represented using model parameters, refining the results based on a single attribute of the item is less trivial. To address this challenge, several works study controllable recommendation via disentanglement~\cite{Cui:2021:disentagled}, where the goal is to represent items as separated attribute vectors instead of a single latent vector. Some of these attributes would be mapped to actual attributes in the catalog (e.g., color, style), or some latent attributes. \Acp{LLM} have shown to be capable of extracting query facets, relying solely on their intrinsic knowledge~\cite{Lee:2024:enhanced}. However, as shown in the literature, \acp{LLM} are not yet capable of effectively grounding~\cite{Shaikh:2024:grounding}, which leads to suboptimal planning of \acp{LLM} utilizing their intrinsic knowledge to take the best next action. For example, for cases where humans would 
As the items or documents are being represented using model parameters, refining the results based on a single attribute of the item is less trivial. To address this challenge, several works study controllable recommendation via disentanglement~\cite{Cui:2021:disentagled}, where the goal is to represent items as separated attribute vectors instead of a single latent vector. Some of these attributes would be mapped to actual attributes in the catalog (e.g., color, style), or some latent attributes. \Acp{LLM} have shown to be capable of extracting query facets, relying solely on their intrinsic knowledge~\cite{Lee:2024:enhanced}. However, as shown in the literature, \acp{LLM} are not yet capable of effectively grounding~\cite{Shaikh:2024:grounding}, which leads to suboptimal planning of \acp{LLM} utilizing their intrinsic knowledge to take the best next action. For example, in conversations where most humans would ask for refinement, \acp{LLM} fail to take the same action.

\textbf{Interactive challenges.}
As mentioned above, there has been research on various modes of refinement, i.e., search verticals, item attributes, faceted search, and example critique. While each of these modes has been utilized for a specific interaction medium (e.g., web search vs.\ conversational search), generative systems could potentially mix them. For example, prompting the user about their preferred search result modality, rather than making an assumption.
Moreover, Chen et al.~\cite{Chen:2023:when} review the interactive challenges of \acp{LLM} in the light of personalization, highlighting the importance of user--system interactions in result presentation, specifically refinement. Among other challenges, they refer to laborious data collection for training \acp{LLM} to be effective interactive systems, which can hinder the learning process.

\section{Clarification}%
\label{clarification}

In a generative retrieval setting where the system aims to provide a comprehensive response to the user, whether in a conversational or web search setting, it is of utmost importance to ensure that the user's intent is predicted with high confidence. This is particularly critical, as in traditional web search scenarios, the system would diversify the list of results to ensure that various facets or interpretations of the query are covered in the top results~\cite{Santos:2015:search}. However, in a generative scenario, usually, a single answer is provided to the user, limiting the information that can be exchanged between the user and the system.

\subsection{An Overview of Search Clarification}

Clarifying questions have been studied extensively~\cite{Keyvan:2023:how} in the context of conversational question-answering~\cite{Rao:2018:learning}, information-seeking conversations~\cite{Aliannejadi:2019:asking}, and web search~\cite{Zamani:2020:generating}. 

Another line of research studies the role of mixed-initiative interactions for user preference elicitation~\cite{Radlinski:2019:coached,Kostric:2024:generating}. The goal here is to understand the user preference when multiple documents (items) can be deemed relevant to their information need. Radlinski et al.~\cite{Radlinski:2019:coached} study this problem for movie recommendation, where the user information need is typically generic (e.g., ``romantic movies'') with multiple potentially relevant items. The dialogue system's goal in this setting is to engage in a conversation to elicit user preference in a more fine-grained way.  

There has been a body of research studying the effect of mixed-initiative interventions such as clarifying questions on user experience~\cite{Kiesel:2018:toward,Zamani:2020:analyzing,Zou:2023:users,Zou:2023:asking}. Kiesel et al.~\cite{Kiesel:2018:toward} study the effect of voice query clarification on user experience and find even in cases where the system performance is not improved, users have better experience. 
In web search, Zamani et al.~\cite{Zamani:2020:analyzing} study the effect of incorporating a clarification pane on the search result page, implemented in Bing.com. Analyzing the click logs, they find that the clarification pane improves user experience. More specifically, among the seven templates they use to generate the clarifying questions, they find clear preference towards certain question templates in terms of user engagement.
Zou et al.~\cite{Zou:2023:users} study the effect of the clarification pane in the same setting in a controlled experimental setup where they introduce three quality levels and measure user satisfaction and performance. They find that asking a low-quality question in a search session risks lower user engagement with questions of higher quality in the same session. This finding was confirmed in a follow-up work~\cite{Zou:2023:asking}.

User engagement (i.e., click-through rate) can be considered as a user-oriented quality measure of clarifying questions. 
Sekulic et al.~\cite{Sekulic:2021:user,Sekulic:2022:exploiting} extract various SERP- and document-based features to predict user engagement while interacting with clarifying questions in a web-based interface~\cite{Zamani:2020:mimics}. Rahmani et al.~\cite{Rahmani:2024:clarifying} study the effect of various query- and question-based features to predict user satisfaction in the MIMICS dataset~\cite{Zamani:2020:mimics} where they find, among others, a positive sentiment in the clarifying question leads to higher user satisfaction. Sekulic et al.~\cite{Sekulic:2024:estimating} instead predicts the usefulness of clarifying questions in the retrieval pipeline. Following an early study on the effect of different types of clarifying questions on retrieval performance~\cite{Karsakis:2020:analysing}, they train a classifier to predict the usefulness of a clarifying question and its answer in the retrieval pipeline and incorporate it in the retrieval pipeline if only it is predicted to be useful.

\subsection{Technical Challenges}

\textbf{Planning.}
While the early works in this area focused mainly on ranking clarifying questions from a pre-collected question bank~\cite{Aliannejadi:2020:convai3,Aliannejadi:2021:building}, more recent studies aim towards leveraging the generation power of \acp{LLM} to generate clarifying questions~\cite{Zhang:2023:ask}. However, generative systems based entirely on \acp{LLM} are not effective in proactive interactions, especially in generating clarifying questions when necessary~\cite{Deng:2024:large,Shaikh:2024:grounding}. Initial experiments reveal the power of \acp{LLM} in understanding the context of a query or a search session~\cite{Abbasiantaeb:2024:let} and generate potential questions based on the context when prompted~\cite{Deng:2023:survey}; however, they fail at planning when to ask and which question to ask~\cite{Deng:2023:survey,Shaikh:2024:grounding}. 
Shaikh et al.~\cite{Shaikh:2024:grounding} conduct a study where they compare human--human conversations with system--human conversations and find that \acp{LLM} fail at effectively planning when to ask clarifying questions in a conversation, even though they can generate high-quality questions if they are explicitly prompted to do so.
Deng et al.~\cite{Deng:2023:prompting} propose a proactive chain-of-thought approach to enhance the planning capability of \acp{LLM} such as ChatGPT and show that it has a considerable effect on their interaction capabilities. 

\textbf{Evaluation.}
Evaluating generative systems comes with various challenges. On top of that, evaluating interactive generative systems involves even more challenges as the user response to a system output is required. 
A line of research looks at simulating and modeling the user--system interactions in a mixed-initiative setting~\cite{Zhang:2020:evaluating,Balog:2021:conversational,Salle:2022:cosearcher,Aliannejadi:2021:analysing,Azzopardi:2022:towards,Owoicho:2023:exploiting,Sekulic:2022:evaluating}. User simulation can be beneficial to generative IR models in two ways: (i) they provide a means for evaluating generated content, and (ii) they can be used for training. Zhang and Balog~\cite{Zhang:2020:evaluating} propose a user simulator for conversational recommendation to evaluate the system performance. This is followed by the work done by Sekulic et al.~\cite{Sekulic:2022:evaluating} and Owoicho et al.~\cite{Owoicho:2023:exploiting} in using GPT-based models to simulate users in a mixed-initiative information-seeking conversational system where the main goal of the simulator is to provide an answer to a generated clarifying question. They show that such simulators can lead to reliable evaluation of conversational systems. 

There are various considerations to take into account in simulating and evaluating interactive generative systems:
\begin{itemize}
    \item User effort: In interacting with the system, users bear different levels of cognitive load, which can lead to user fatigue as the number of interactions increases.
    \item User information gain: To model the true value of a clarifying question in a conversation, we need to model both the gain and effort a clarifying question brings to the conversation~\citep{Aliannejadi:2021:analysing,Azzopardi:2022:towards}. 
    \item Information nuggets: Information gain can be modeled by breaking the user's information need into information nuggets and measuring how much asking a certain clarifying question would help us provide further information nuggets to the user.
    \item User model: As proposed by Balog~\cite{Balog:2021:conversational}, an effective user simulator should have various components, including a user mental model. Realistically, a single user simulator does not cover the needs and behavior of the wide range of users interacting with the system. 
\end{itemize}

\section{Proactive Interactions}%
\label{proactive}
Typically, users initiate the interaction with a generative retrieval system, for example by submitting a chit-chat utterance, asking a question, or submitting an action request. In mixed-initiative conversational systems, the agent is also able to initiate the conversation. This is also called a \emph{proactive}, system-initiative, or agent-initiative conversation. Existing generative AI systems are relatively under-developed when it comes to proactive interactions \cite{Liao:2023:proactive}. A major reason is that initiating a conversation by the system is not only challenging, but can also be risky; frequent and non-relevant proactive interactions may annoy users and hurt user satisfaction and trust \cite{CISbook}. Therefore, whether and when to initiate a proactive interaction are the key decisions a proactive CIS system should make.

\begin{figure*}[t]
    \centering
    \includegraphics[width=\textwidth]{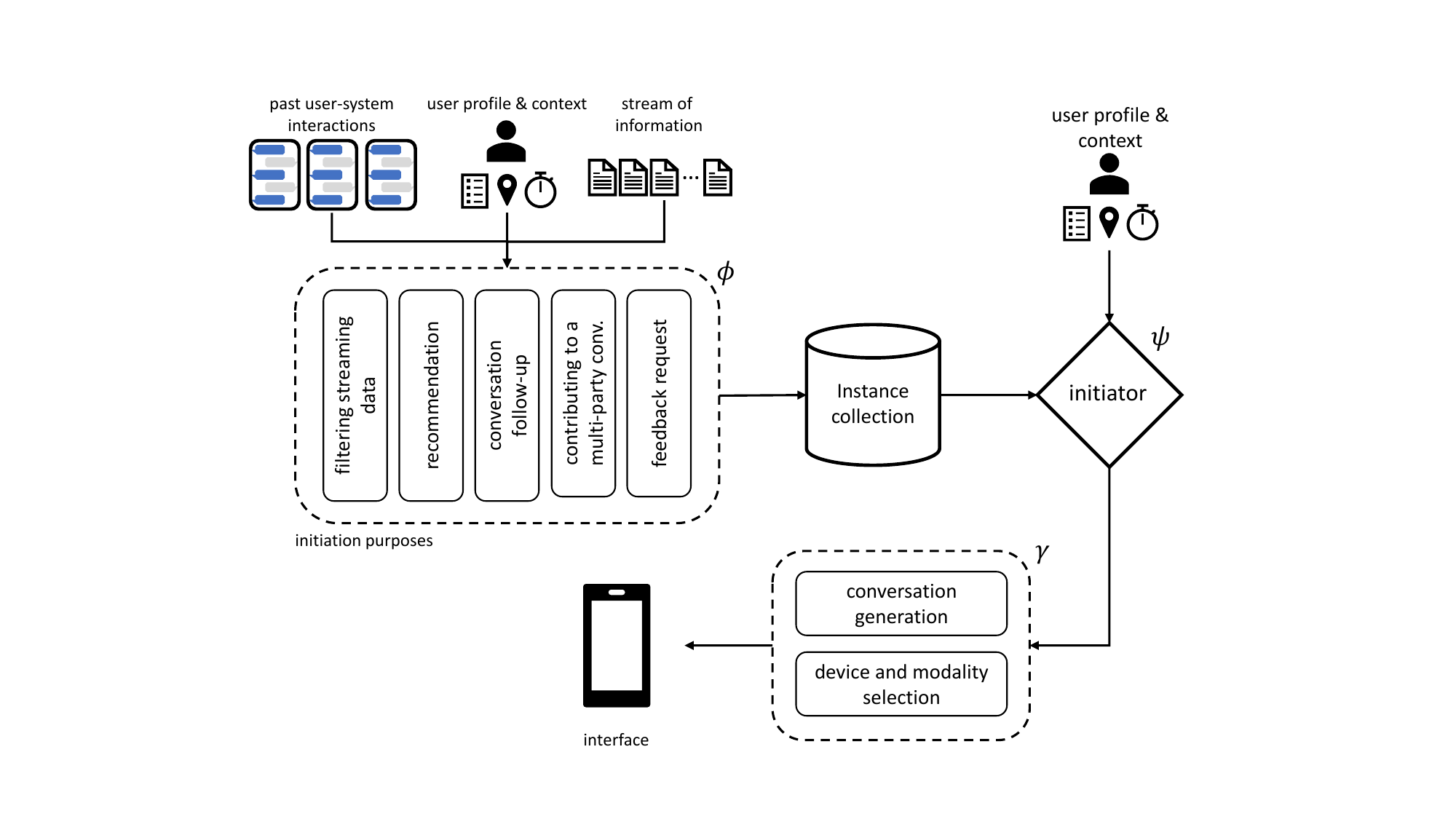}
    \caption{A generic pipeline for conversation initiation in CIS systems by Wadhwa and Zamani~\cite{Wadhwa:2021:SystemInitiative}.}
    \label{fig:proactive}
\end{figure*}

\subsection{An Overview of Proactive Generative Retrieval Systems}
Wadhwa and Zamani~\cite{Wadhwa:2021:SystemInitiative}  explored proactive conversational information access systems, discussing their challenges and opportunities.  The authors introduced a taxonomy of proactive interactions, delineating three dimensions: (1) initiation moment (\emph{when} to initiate a conversation), (2) initiation purpose (\emph{why} to initiate a conversation), and (3) initiation means (\emph{how} to initiative a conversation). They identified five purposes for initiating interactions:  (1) filtering streaming information, (2) context-aware recommendation, (3) following up a past user-system conversation, (4) contributing to a multi-party human conversation, and (5) requesting feedback from users. A generic pipeline for these systems is depicted in Figure~\ref{fig:proactive}. In this pipeline, several algorithms constantly monitor the user's context and information streams to produce conversation initiation instances, which are stored in a database. A conversation initiator component then selects an appropriate instance based on the situation, initiating a fluent and accurate utterance. Figure~\ref{fig:proactive} is sufficiently generic for illustrating proactive interactions in generative retrieval models and we use it to describe research and open questions in proactive retrieval in more detail.

Initiating a conversation through recommendation stands as one of the most common scenarios for proactive interaction. For instance, a conversational information access system might suggest an item based on the user's situational context, such as their location, time, and preferences. It is worth noting the distinction from traditional conversational recommendation setups, where users typically initiate the conversation by requesting specific items \citep{Sun:2018:ConvRecSys,Zhang:2018:cikm}. Recent efforts in joint modeling of search and recommendation and developing unified information access systems \citep{Zamani:2018:DESIRES,Zamani:2020:JSR,Zeng:2023:UIA} represent a step towards developing proactive, and thus mixed-initiative, systems in search and recommendation. However, proactive conversations extend beyond mere recommendations.

For example, Avula and Arguello~\cite{Avula:2020:Wizard} devised a system for conducting wizard-of-oz experiments, investigating proactive interactions during conversational collaborative search. This system could seamlessly integrate into collaborative platforms like Slack,\footnote{\url{https://slack.com/}} where during a collaborative search task, an external u    ser (acting as a wizard) provides information. Though advancements in this area are nascent, there exists considerable potential for systems to initiate context-based conversations, engaging users and eliciting feedback.

Consider a scenario where a user employs a mapping application to navigate to a restaurant. Leveraging contextual cues, a proactive generative retrieval system could subsequently initiate a conversation upon the user's return journey, inquiring about their dining experience. Such interactions not only enhance user engagement but also facilitate feedback collection, aiding in profile refinement. Similarly, in situations where a user encounters difficulty in task completion, a conversational system could autonomously engage in conversation, offering assistance \cite{CISbook}.

\subsection{User Responses to Proactive Interactions}
While in generative retrieval systems, users have the freedom to provide a natural language response in any form, they can be categorized as follows \cite{Wadhwa:2021:SystemInitiative}:
\begin{itemize}
    \item Null action: Users provide no response to the initiated conversation. It is important to note that null action should not necessarily be construed as negative feedback, as users may find the initiation useful but may not desire further engagement. 
    \item Interruption or negation: Users respond in a manner consistent with terminating any further engagement by the generative retrieval system. It is perhaps safe to interpret such responses as negative feedback.
    \item Relevant response: Users provide a pertinent response to the initiated interaction, typically occurring when the interaction involves a question or solicits feedback.
    \item Postpone: Users respond to the initiated conversation and request the system to remind them at a later time.
    \item Critique or clarification-seeking response: Users engage further with the generative retrieval system, either seeking more information or critiquing existing engagement. 
    \item Follow-up: Users provide a follow-up response to obtain additional information or perform actions related to the initiated conversation.
    \item Topic drift: Users respond but shift the topic of the initiated conversation.
\end{itemize}

\subsection{Technical Challenges}
Here, we outline key technical hurdles in implementing the pipeline shown in Figure~\ref{fig:proactive}.

\textbf{Producing System-Initiative Instances.}
The initial step in the system-initiation pipeline involves identifying reasons for initiating a conversation and generating a proactive instance. Proactive instances encapsulate all relevant information about a conversation, including its purpose, content, and context. This process entails addressing each initiation purpose component outlined in Figure~\ref{fig:proactive}. While some purposes, such as filtering streaming information and recommendation, have received attention in the literature, others like following up a past conversation or contributing to a multi-party conversation remain relatively unexplored. Thus, a major technical challenge lies in developing models capable of identifying the reasons for conversation initiation across various goals, including filtering information, recommendation, conversation follow-up, contributing to multi-party conversations, or requesting feedback.

\textbf{Developing an Initiator Model.}
The subsequent step involves selecting a proactive instance from the instance collection using an initiator component. The primary challenge in this component stems from our limited understanding of the optimal moment to initiate a conversation. Consequently, future research should emphasize conducting user studies to explore the ideal timing for conversation initiation. Weak signals gleaned from user interactions with existing conversational systems, even those lacking proactive capabilities, could provide valuable insights. For instance, instances, where users initiate trivial conversations (e.g., out of boredom), could serve as noisy but potentially useful signals for predicting optimal conversation initiation moments. Machine learning models trained on situational context and user profiles could leverage such signals. Furthermore, interactive systems that log user interactions offer the opportunity to iteratively refine prediction accuracy based on user feedback.

\textbf{Generating System-Initiative Utterances.}
The final step entails generating a (natural language) interaction based on a proactive instance and presenting it to the user. Techniques from dialogue systems and text generation research can be leveraged for this purpose. Since users typically do not anticipate proactive utterances, a notable technical challenge lies in providing context within the generated utterance to ensure user comprehension. This context could reference previous interactions with the system, user experiences, or explanations regarding the rationale behind initiating the conversation. Given that each instance is a structured data object, neural models designed for unstructured text generation from structured data, such as tables, could be potentially useful.

\subsection{Evaluation of Proactive Systems}
Assessing proactive generative IR systems poses significant challenges. While IR research has traditionally focused on creating collections for specific information-seeking tasks, these collections are typically based on predefined needs (e.g., TREC\footnote{\url{https://trec.nist.gov/}} tracks) or observations (e.g., clickthrough data). However, these evaluation methods do not readily apply to scenarios involving proactive interactions. Although evaluating proactive generative IR systems remains largely unexplored in the literature, we can envision two classes of evaluation methodologies: (1) modular evaluation, and (2) end-to-end evaluation.

In modular evaluation, the quality of each component in Figure~\ref{fig:proactive} is evaluated in isolation? For example, how accurate is the initiator component in identifying opportune moments for proactive interactions? This  methodology simplifies evaluation in proactive systems, but does not provide a complete picture of the overall performance of the system from the user's perspective, and does not reflect real-world complexities.

In end-to-end evaluation, one can explore both offline and online evaluation strategies. For offline evaluation, each instance would encompass all necessary information for the system at a given timestamp, including past user-system interactions, user profiles, situational contexts, and streams of new information. The model's performance would then be assessed based on the generated proactive interactions, if applicable. Crafting a single evaluation metric capable of capturing all facets of conversation initiation evaluation presents a challenge, necessitating further investigation. Recently, Samarinas and Zamani~\cite{Samarinas:2024:ProCIS} introduced a large-scale benchmark for proactive interactions to ongoing multi-party human conversations and proposed normalized proactive discounted cumulative gain (npDCG) for end-to-end evaluation of such systems. In a separate investigation, Sen et al.~\cite{Sen:2018} suggested evaluating proactive recommendation within search sessions by aggregating a correlation measure over the session. This measure assesses the relationship between the expected outcome---comprising the list of documents retrieved with a true user query---and the predicted outcome, representing the list of documents recommended by a proactive search system.

In the realm of online evaluation, conventional A/B tests can serve as a valuable tool for assessing the system's efficacy. Additionally, interpreting user feedback---both positive and negative---can provide valuable insights into system performance.

\section{Explanation}
\label{explanation}
Explanation can be seen as a critical tool in search result presentation in generative systems, as users are interested in comprehensive justification and explanation of the presented results~\cite{Gao:2019:explainable, Chen:2023:when}. Also, it can lead to more user trust in the results, potentially aiding the user to distinguish between a low-quality and a high-quality response.

\subsection{An Overview of Explanation in IR}
Zhao et al.~\cite{Zhao:2023:Explainability} provide a survey on the explainability of \acp{LLM} where they provide a taxonomy of explanations, together with methods for explaining Transformer-based \acp{LLM}. Also, they discuss various methods for evaluating explanations for both local and global explanations. 
Krishna et al.~\cite{Krishna:2023:posthoc} show that not only are explanations useful in user--system interactions, but they also improve the performance of \acp{LLM}. They study automatic rationale generation in a \ac{CoT} manner.
Deng et al.~\cite{Deng:2023:rephraserespond} show that rephrasing the user input leads to a better understanding of the user request which in turn results in better performance of the LLM, which is complementary to \ac{CoT} reasoning. 
In their tutorial, Anand et al.~\cite{Anand:2022:explainable,Anand:2023:explainable} review Transformer-based explanation generation. 
Zhang et al.~\cite{Zhang:2020:query} addresses search explainability via the lens of query understanding, where the system's task is to predict the user intent considering their query as input. 
LiEGe~\cite{Yu:2022:towards} explains all the documents in the ranking jointly using a listwise explanation generator. 

Evaluating explanations is challenging. For free-text generations, human evaluation is employed. In other cases, because of a lack of explanation, proxy explanations such as clicks, query descriptions, query aspect annotation, and topic annotation can be used. For feature-based models, explanations are evaluated based on the effectiveness of predicted features. As for counterfactual explanations, model-based evaluation is employed.

\subsection{Modes of Explanation in Generative IR}
The main mode of explanation used in generative models is free-form text, where the model would further elaborate why the provided answer is relevant to the user's input. The explanation often consists of two major parts: (i) a further description of user information need, and (ii) an explanation of the reasons why the generated response is relevant to the user's input.
The system has a limited information bandwidth and cannot present the users with multiple intents of their query. Therefore, describing what the system ``thinks'' the user wants helps the user understand whether the system understands their intent or not~\cite{Zhang:2020:query}.
This type of explanation aims to ensure the user that their information need is properly understood by the system and can lead to increased trust in the system. Also, in case of misunderstanding the user's information need, it provides the opportunity for the user to realize what is missing in their input. This can be seen as similar to scanning the SERP by the user, through which the user would have an idea if the system understands their information need correctly.

Another form of explanation is to provide citations. This has been studied more extensively in the NLP community where the generated text attribution \citep{Huang:2023:citation}. The URL citations are supposed to provide evidence of the source of information from the web. However, there are concerns regarding the quality of the citations, as there is no clear way of controlling the \ac{LLM} to ground its responses on the cited page~\citep{Zhang:2024:towards}. Citing source documents, while being useful as a form of explanation, still does not provide a comprehensive idea of the relevance of the source. Comparing it to a typical SERP where the users are exposed to the URLs of the results, users already have a quality perception by scanning through the page title, summary, and URL. Even though the \ac{LLM}-based search interfaces aim to mimic this experience, it is not yet clear which parts of the generated response are extracted from the cited document. Moreover, it is not clear how much the system depends on its intrinsic knowledge (i.e., model parameters) vs.\ the retrieved document. Therefore, more research in this area is required to understand how much different techniques and modes of explanation affect the users' perception of quality and trust. One potential alternative is to treat the system as an information-gathering tool~\cite{Ren:2021:wizard}, rather than an information system. In such cases, the responses would look like  ``After searching the web, I found numerous sources of information about your query. Two of more trustworthy sources mention that ....'' With such a response, not only does the user learn about the search space of the given query, but also they learn about the most important information extracted from the topic documents.

\section{Multi-Modal Interactions} \label{multimodal}
Research has demonstrated the advantageous role of multimodal signals in both keyword-based and recommendation-driven searches, spanning from contextual item recommendations \cite{DBLP:conf/recsys/KaminskasRS13,DBLP:conf/www/WangWTSRL17} to visual and multimedia recommendations \cite{DBLP:conf/sigir/McAuleyTSH15}. These signals also address challenges like cold-start issues \cite{DBLP:journals/umuai/DeldjooDCECSIC19,DBLP:conf/recsys/OramasNSS17,DBLP:conf/mm/CuiWS14} and aiding in explaining and visualizing recommendation outcomes \cite{DBLP:conf/wacv/TangsengO20}. A recent survey by Deldjoo et al.~\cite{DBLP:journals/csur/DeldjooSCP20} offers insights into the role of multimedia content in recommendation systems, delineating how such content—comprising audio, visual, and textual elements—enriches real-world recommendation challenges. 

A significant challenge in Multimedia Information Systems lies in fusing multiple modalities to derive meaningful representations. Recent advancements in multi-modal large language models employ joint representation techniques to establish a latent space where multiple modality information can be compared. However, aligning content data like text and images is relatively straightforward compared to aligning content with user preferences such as ratings or social media data.

Deldjoo et al.~\cite{Deldjoo:2021:MMCIS} explored multi-modal conversational information-seeking tasks from multiple perspectives. They investigated (1) \textit{Why} using multi-modal interactions, (2) \textit{Which} tasks to support in multi-modal conversational systems, (3) \textit{When} to integrate multiple modalities in conversations, and (4) \textit{How} to research multiple modalities and conversations to enable multi-modal conversational information seeking. Deldjoo et al.~\cite{Deldjoo:2021:MMCIS}  highlight the importance of each of these perspectives through a real-world example: 

\begin{quote}
Imagine a person is cycling along the road on their way to work. She is planning her day, including tasks from presenting a budget, hosting a new client, picking up their children after school, and making dinner. The cyclist passes a flower on the sideroad, which caught her eye and wanted to know what this plant is. Since she is cycling on a busy road, she quickly stops, takes a photo, and keeps riding. Meanwhile, she asks her earbuds to tell her which plant that was by a spoken query such as ``what was that plant and is it edible?''
\end{quote}

The authors argue that generative IR systems with multi-modal interactions and multi-modal sensors can accomplish the user's need in this and even more complex scenarios.
Dealing with multi-modal interactions is a multidisciplinary topic, spanning across research areas from information retrieval, recommender systems, multi-media, human-computer interactions, computer vision, and even psychological and cognitive sciences. The intersection of the research areas that enable people to search for information through multi-modal conversations has not received the attention it deserves and it might partially be due to the complexity of the topic in terms of both modeling and evaluation. Prior work are mostly limited to two modalities (image and text), e.g., \cite{sundar-heck-2022-multimodal,10.1145/3543829.3543837}, and further development in multi-modal foundation models  \cite{CGV-110,DBLP:journals/corr/abs-2110-14378} and multi-modal retrieval-augmented generation models \cite{10.1145/3539618.3591629}, is expected to speed up progress in this area.

\section{User Interfaces}
\label{interface}

While in Section  \ref{infoneeds} we focused on general interaction methods to assist users in expressing their information needs when interacting with generative AI, in this section we  review recent work on interaction techniques and user interfaces for information access with \acp{LLM}. The design space is huge, and it is still under-researched and poorly understood. For example, out of approximately 750 pre-prints related to \acp{LLM} published on arXiv in the field of Information Retrieval between 2020-2024, only 22 mentioned ''user interface'' in their abstracts. 

New human-LLM interaction frameworks are only starting to emerge. For example, recent work \cite{gao2024TaxonomyHumanLLMInteraction} reviewed 73 papers published in HCI conferences  since 2021 to investigate the dynamics of human-LLM interaction. Authors identified four key phases in the interaction flow and developed a taxonomy of four primary interaction modes. The four phases are: \textit{planning} - before an interaction, \textit{facilitating} - during an interaction, \textit{iterating} - refining an interaction, and \textit{testing} - testing an interaction. The interaction modes include: \textit{Standard Prompting, User Interface, Context-based and Agent Facilitator}. The \textit{User Interface} mode is of most interest to us as it enhances user interactions with \acp{LLM} beyond the conversational interface by improving input, output, iteration, and reasoning processes. This mode contains five approaches, which could be used separately or in combination. 
(1) \textit{Structured prompt} approaches assist users in creating multi-component prompts, which could range from zero-shot to few-shot, and support specification of constraints. Tools like PromptMaker \cite{jiang2022PromptMakerPromptbasedPrototypinga} combine prefixes, settings, and examples in prompt creation. 
(2) \textit{Varying output} approaches allow users to specify output formats. Early examples like GenLine and GenForm \cite{jiang2021GenLineGenFormTwoa} facilitate generation of user specified mixed outputs, such as HTML, JavaScript, and CSS code. User's control over output format allows for high level of personalization and, potentially, enhances consumption of information.
(3) \textit{Iteration of interaction} approaches include features such as debugging, error labeling, regenerating, and self-repairing, enabling users to refine their original prompts and workflows. BotDesigner \cite{zamfirescu-pereira2023WhyJohnnyCana}, for instance, helps users identify and label errors within conversations and offers a ''retry'' button to regenerate outputs.
(4) \textit{Testing of interaction} facilitates the testing of various prompt variations, useful for quick testing of complex solutions. Tools like VISAR \cite{zhang2023VISARHumanAIArgumentative} use visual programming to enable rapid prototyping and testing of writing organization.
(5) \textit{UI to support reasoning} incorporates direct manipulation in the Chain-of-Thought process, allowing users to actively participate in and reorganize reasoning sequences. Other approaches in this area offer visual programming techniques, such as chain designs and mind maps and enable a more interactive and user-defined reasoning framework \cite{jiang2023GraphologueExploringLarge, suh2023SensecapeEnablingMultilevel, zhang2023VISARHumanAIArgumentative}. For example, Graphologue \cite{jiang2023GraphologueExploringLarge} introduced: (1) graphical diagrams which convert text-based responses from \acp{LLM} into diagrams; (2) graphical dialogues which enable graphical, non-linear dialogues between humans and \acp{LLM}; and (3) interactive diagrams which allow users to adjust graphical presentation, its complexity and submit context-specific prompts.

MacNeil et al.~\cite{macneil2023PromptMiddlewareMapping} explores three methods for integrating \acp{LLM} into user interfaces through a framework called Prompt Middleware.  The three methods are: (1) \textit{Static Prompts} are predefined prompts generated by experts through prompt engineering. They can be invoked by using UI elements (e.g., buttons), allowing users to send high-quality prompts to with minimal effort. This method leverages best practices but limits user control over prompt generation. (2) \textit{Template-Based Prompts} involve generating prompts by filling in a template with options selected from the UI. The template integrates expertise and best practices, giving users more control through UI options. This method is exemplified by the FeedbackBuffet prototype, a writing assistant that uses template-based prompts to generate feedback on writing samples \cite{macneil2023PromptMiddlewareMapping}. (3) \textit{Free-Form Prompts}: This method grants users full control over the prompting process. Although challenging, it is beneficial in scenarios where complete control is desired.

Wang et al.~\cite{wang2024TaskSupportivePersonalizeda}
present a proactive interface design that addresses challenges users face in initializing and refining prompts, providing feedback to the system, and managing cognitive load. They describe three interaction techniques (\textit{Perception Articulation, Prompt Suggestions, Conversation Explanation}) and how they can be supported by user interface elements. Perception articulation is supported by a pre-task questionnaire and main prompt template - the first supports information need at the visceral level, while the latter at the formalized level. Prompt suggestions are provided through supportive function tabs, which support conscious need. Conversation explanations are also delivered through supportive function tabs, with a feedback mechanism allowing users to rate the usefulness of these explanations. This feature supports compromised needs.
Evaluation with participants demonstrated the effectiveness of these supportive functions in reducing cognitive load, guiding prompt refinement, and increasing user engagement. In interviews, participants appreciated the perception articulation functions for setting expectations and the conversation explanations for balancing expectations and satisfaction.

On one hand, the design space of user interfaces for \acp{LLM} offers a myriad of new interaction possibilities. On the other, taking advantage of the new possibilities can lead to complexity, which can make interfaces harder to comprehend and can overwhelm users. From the history of search interface evolution, we know that more complex search interfaces have not been widely accepted. For example, faceted search UIs led to a sharp learning curve and increased cognitive load \cite{wilson2011EvaluatingCognitiveImpact}. History likes to repeat itself. Will it be the case with user interfaces for \acp{LLM}? Will the more complex interfaces for \acp{LLM} become only niche products? 

\section{Conclusions} \label{conclusions}
As mentioned multiple times throughout this chapter, handling complex interaction types and modalities has been relatively under-explored and the authors find it a rich area of investment for the further development of generative information retrieval systems. This chapter pointed out prior work on various interaction types, from expressing information need to result refinement and mixed-initiative interactions, including clarification, feedback, and proactive interactions. Recent developments in (multi-modal) foundation models, including large language models, have paved the path towards better understanding complex user interactions, but we are still far from ideal generative information retrieval systems that can satisfy user needs efficiently, effectively, fairly, and robustly.

\bibliographystyle{splncs04}
\bibliography{XX-references}

\end{document}